# System Identification via Validation and Adaptation for Model Updating Applied to a Nonlinear Cantilever Beam

**Cristian López**
Department of Mechanical and Materials Engineering, University of Nebraska-Lincoln
W342 Nebraska Hall, Lincoln, NE 68588
clopez19@huskers.unl.edu

**Jackson E. Herzlieb**
Daniel Guggenheim School of Aerospace Engineering, Georgia Institute of Technology
270 Ferst Dr NW, Atlanta, GA 30332
jherzlieb3@gatech.edu

**Keegan J. Moore[1]**
Daniel Guggenheim School of Aerospace Engineering, Georgia Institute of Technology
270 Ferst Dr NW, Atlanta, GA 30332
kmoore336@gatech.edu
ASME Member

**ABSTRACT**

*The recently proposed System Identification via Validation and Adaptation (SIVA) method allows system identification, uncertainty quantification, and model validation directly from data. Inspired by generative modeling, SIVA employs a neural network that converts random noise to physically meaningful parameters. The known equation of motion utilizes these parameters to generate fake accelerations, which are compared to real training data using a mean square error loss. For concurrent parameter validation, independent datasets are passed through the model, and the resulting signals are classified as real or fake by a discriminator network, which guides the parameter-generator network. In this work, we apply SIVA to simulated vibration data from a cantilever beam that contains a lumped mass and a*

---

[1] Corresponding author

*nonlinear end attachment, demonstrating accurate parameter estimation and model updating on complex, highly nonlinear systems.*

## 1. INTRODUCTION

The longevity and optimal operation of structures and devices depend on the ability to understand and manage vibrations [1,2]. In structural dynamics, SI plays a key role, e.g., by using vibration data, to obtain mathematical models that describe the system response, estimate parameters such as mass, stiffness, and damping, and predict dynamic behavior [3,4]. SI is essential across various fields. In structural health monitoring, it enables real-time damage detection and assessment of load-carrying capacity [5]. In vibration control, precise models help mitigate the effects of external forces, guaranteeing stability during events such as earthquakes or strong winds [6]. In noise control, accurate system modeling offers effective noise reduction strategies [7], etc.

Historically, vibration analysis methods have depended on theoretical models that are validated by experimental testing specifically created for those models [8]. However, techniques grounded in linear assumptions and Fourier analysis frequently fall short in representing the nonlinear and nonstationary dynamics of many systems [9]. In contrast, contemporary data-driven approaches enable the direct characterization of structural dynamics from measured data [10], presenting new possibilities to tackle complex behaviors [11,12].

Advances in nonlinear SI in structural systems have been reviewed in Kerschen et al. [3] and Noël et. al [4]. SI methods can be categorized into three groups:

*parametric methods* rely on a known mathematical model, where the identification process concentrates on determining its coefficients. These include moving average models [13, 14], Kalman filter [15], Bayesian methods [16,17], nonlinear state-space system models [18], time-series methods [19,20], moving horizon optimization [21], as well as integrating physics with neural networks (NNs) [22–27], etc.; *non-parametric methods* infer the system's dynamics from data without assuming a prior model, such as the restoring force surface method [28,29], using genetic programming [30–32], NNs [33–37], slow-flow models [38,39], the Nonlinear Identification through eXtended Outputs method [40], combining symbolic regression and genetic programming [41], a data-driven based machine learning and symbolic regression approach [42], among others; *semi-parametric methods* incorporate measured data with partial knowledge of the system to identify the unknown dynamics. Examples encompass the piecewise-linear RFS method [43], the sparse identification of nonlinear dynamics (SINDy) method and its Bayesian extension [12,44,45], characteristic nonlinear SI method [46], Hamiltonian-constrained autoencoder [47], and energy-based methods [48,49].

Recent advances in artificial intelligence have led to the widespread adoption of data-driven methods for SI. Cunha et al. [50] reviewed computational intelligence approaches for nonlinear SI, and Quaranta et al. [51] surveyed machine learning applications in structural dynamics and vibroacoustic. These frameworks leverage machine learning's ability to discover complex patterns, reducing the dependency on physics-derived models. Various architectures have been applied to SI: feed-forward NN to obtain restoring forces in Duffing oscillators [33], wavelet NNs for structural SI [34], recurrent NN for impact forces estimation [35], convolutional NN for structural dynamic response and SI [36], and symbolic NN to derive governing equations [37]. Including known physics can further improve SI by guiding the model with physical constraints, established parameters and/or governing equations. In this sense, researchers applied long short-term memory network for predicting nonlinear structural responses [22] and hysteretic parameter identification [23]; combined physics-informed NNs and the Runge-Kutta scheme for parameter estimation and dynamical systems modeling [24], applied generative adversarial networks (GANs) for structural parameter identification [25], employed fully connected networks for structural response prediction [26], incorporated Hamiltonian mechanics in an autoencoder for structural dynamics modeling [47], etc.

From these various architectures, GANs stand out as an interesting technique for realistic data generation [52], using an adversarial process between two networks. During training, the generator aims to create plausible synthetic data, while the discriminator learns to distinguish authentic from synthetic data. Applications include content generation (images, text, music) [53–55], stochastic differential equations solving [56], mechanical systems analysis [57,58], etc. In structural mechanics, GANs have been used for model updating [59], parameterized nonlinear systems modeling [60], nonlinear modal analysis [61], and parameter identification [25].

Recently, the so-called System Identification via Validation and Adaptation (SIVA) method [27] was introduced, drawing inspiration from GANs and incorporating physics knowledge in the form of governing equations of motion. By using the displacements and velocities to generate the corresponding accelerations, SIVA employs an adversarial framework to learn the physical parameters that govern the system dynamics. Uniquely, it not only performs data-driven and physics-based parameter identification, but also uncertainty quantification and model validation simultaneously. In this paper, we apply the SIVA method to the analysis and model updating of a beam with a strongly nonlinear end attachment.

## 2. THE SYSTEM IDENTIFICATION VIA VALIDATION AND ADAPTATION (SIVA) METHOD

In the following, we provide a brief description of the SIVA method, which is illustrated in Fig. 1. Throughout this paper, we use bold lowercase letters for vectors and bold capital letters for matrices. The SIVA approach calculates the parameters of the system using training acceleration time series while concurrently validating them with unseen accelerations. This is achieved by iteratively updating the system's parameters to match the predicted accelerations and training data. The estimated parameters are also used to generate accelerations from the validation dataset, which are evaluated by a discriminator network to perform validation. Thus, the SIVA method estimates and validates the parameters of the differential equation directly from measured vibrations. Furthermore, after convergence is achieved, uncertainty quantification is conducted on the estimated parameters.

### 2.1. Methodology

For this study, we assume that a linear finite element (FE) model has been identified and updated to match linear test data, and the user now wants to update that existing FE model to incorporate additional linearities and nonlinearities introduced by modifying the system. In the present paper, a cantilever beam serves as the underlying linear structure with an identified model, and the nonlinear end attachment represents the modification that needs to be identified. A reduced-order model for the linear FE model is then constructed using either Guyan reduction [62], the system equivalent reduction expansion process (SEREP) [63], or similar technique as chosen by the user. The reduced-order model is constructed such that the retained DOFs correspond to coordinates that are measured in the experiments and provides the user with the linear mass ($\boldsymbol{M}_R$), damping ($\boldsymbol{C}_R$), and stiffness ($\boldsymbol{K}_R$) matrices that are used in SIVA. In our case, the reduction is applied to retain a subset of only displacement coordinates along the beam. Additionally, the user must have transient, free responses of the system in the form of accelerations, velocities, and displacements at locations that correspond to nodes in the linear model. However, the parameters of the addition to the system (i.e., the nonlinear end attachment) are unknown and need to be identified. In the following, we describe the core steps of the method with specifics included for the attachment identification of the beam simulated in this work:

1. **Data and mathematical model.** Time series data is obtained for the system in the nonlinear configuration. For this paper, this consists of simulating the response of the beam using the full FE model to impact excitations with varying amplitudes. Next, a reduced-order mathematical model $\boldsymbol{f}(\boldsymbol{q}, \dot{\boldsymbol{q}}, \boldsymbol{M}_R, \boldsymbol{C}_R, \boldsymbol{K}_R, \boldsymbol{\kappa}_{nl}; \boldsymbol{\lambda})$ is proposed, where $\boldsymbol{q}$ and $\dot{\boldsymbol{q}}$ denote the displacement and velocity vectors, respectively, $\boldsymbol{\kappa}_{nl}$ denotes the model nonlinear force vector, and $\boldsymbol{\lambda}$ denotes the unknown parameters representing the addition to the system.

2. **Identification of unknown parameters.** The parameter-generator network ($\boldsymbol{P}$) converts a batch of independent random noise $\boldsymbol{z}$ into significant parameter values $\boldsymbol{\lambda}$, which are subsequently used in the proposed reduced-order model, $\boldsymbol{f}$.

3. **Optimization.** The model parameters are updated by minimizing the mean square error (MSE) between real acceleration data $\ddot{\boldsymbol{q}}_{tr}$ and those generated by the model $\widetilde{\ddot{\boldsymbol{q}}}_{tr} = \boldsymbol{f}(\boldsymbol{q}_{tr}, \dot{\boldsymbol{q}}_{tr} \boldsymbol{M}_R, \boldsymbol{C}_R, \boldsymbol{K}_R, \boldsymbol{\kappa}_{nl}; \boldsymbol{\lambda})$; and the adversarial loss, which assesses how well the generated parameters, and consequently the generated accelerations, deceive the discriminator network ($D$). When the discriminator network is sufficiently deceived, the discriminator classifies $\widetilde{\ddot{\boldsymbol{q}}}_{val}$ as real. Thus, using the validation dataset, this loss offers feedback to the parameter generator network indirectly.

4. **Model validation.** Using an independent dataset ($\dot{\boldsymbol{q}}_{val}, \boldsymbol{q}_{val}$) collected under different conditions from those used for training, the model outputs $\widetilde{\ddot{\boldsymbol{q}}}_{val}$, which are assessed by a discriminator network. The discriminator classifies the output as real or fake, and this feedback is used to update both the discriminator and the parameter-generator network. Through the adversarial training, both networks progressively improve, resulting in a framework with built-in validation.

5. **Uncertainty quantification (UQ).** Once convergence is reached, UQ is carried out on the identified parameters either by continuing training and collecting parameter values from additional epochs or by using the trained parameter generator with different batches of noise data without further training. The first approach leverages the framework's robustness to overtraining, while the second approach enables performing UQ immediately after convergence.

## 2.2. Data Processing

In this work, we use an FE model for a cantilever beam with parameters based on geometric and material properties, viscous damping computed using damping ratios and incorporating a local attachment with linear and nonlinear stiffness and a lumped mass. The resulting model is then converted into a state-space form to simulate the nonlinear vibration response of the beam over time. As mentioned in [27], if noise is present in the signals, the SIVA method might struggle to precisely determine a suitable model for the system or produce a reliable acceleration time series. This is especially true when reduced-order matrices produced by Guyan or SEREP are applied directly to experimental data without sufficient smoothing through methods such as [64]. The effects of noise on this approach will be tackled in future research. Since we aim to use translational degrees-of-freedom (DOFs) due to their ease of measurement experimentally, we perform apply Guyan reduction [62] to the mass, damping, and stiffness matrices of the full FE model to retain only the translational DOFs. This reduction results in the reduced mass, damping, and stiffness matrices of $\boldsymbol{M}_R$, $\boldsymbol{C}_R$, and $\boldsymbol{K}_R$, respectively, which are used in the SIVA.

## 2.3. Network Architecture

The methodology has two neural networks: first; a parameter-generator network that converts random noise $\boldsymbol{z}$ to physically significant system parameters $\boldsymbol{\lambda}$; and second, a discriminator network that classifies between real and generated acceleration time series. The details of these networks are presented in Table 1. This framework was implemented in Python 3.12.7 with Pytorch 2.6.0 platform. We use Adam as the optimizer with a learning rate of $10^{-4}$ and a batch size of 500.

## 2.4. Physics-based Modeling

For Newton's second law, for a multiple-degree-of-freedom system, we obtain the fake accelerations $\widetilde{\ddot{\boldsymbol{q}}}$, for both the training and validation datasets, based on the known reduced mass $\boldsymbol{M}_R$, $\boldsymbol{C}_R$, $\boldsymbol{K}_R$ matrices, estimated parameters, and state variables ($\boldsymbol{q}$ and $\dot{\boldsymbol{q}}$) as

$$\widetilde{\ddot{\boldsymbol{q}}} = \boldsymbol{M}_R^{-1}(-\boldsymbol{C}_R\dot{\boldsymbol{q}} - \boldsymbol{K}_R\boldsymbol{q} - \boldsymbol{\kappa}_{nl}(\boldsymbol{q}) + \boldsymbol{F}(t)), \qquad (1)$$

$\boldsymbol{F}(t)$ denotes the external force vector and $\boldsymbol{\kappa}_{nl}(\boldsymbol{q})$ is the nonlinear element to be identified. Since the current work considers model updating with a known source of additional linear and nonlinear stiffnesses, it is natural to only include additional terms at that location. Thus, we define $\boldsymbol{\kappa}_{nl}(\boldsymbol{q})$ as

$$\boldsymbol{\kappa}_{nl}(\boldsymbol{q}) = \left\{ \begin{array}{c} 0 \\ \vdots \\ \sum_{i=1}^{5} k_i q_N^i \end{array} \right\}, \qquad (2)$$

such that the nonlinear element acts at the tip (or the $N$th DOF in this case) of the beam only. However, if the user is unaware of the source of the additional terms, then a general approach can be implemented by populating $\boldsymbol{K}_{nl}(\boldsymbol{q})$ with as many terms in as many DOFs as desired.

### 2.4.1 Loss Functions

In the SIVA method, the networks are trained using standard adversarial learning [52]. The discriminator loss utilizes the binary cross-entropy criterion:

$$\mathcal{L}_D = -\mathbb{E}_{\ddot{\boldsymbol{q}} \sim p(\ddot{\boldsymbol{q}})}\big[\log\big(D(\ddot{\boldsymbol{q}})\big)\big] - \mathbb{E}_{\boldsymbol{z} \sim p(\boldsymbol{z})}\Big[\log\Big(1 - D(\widetilde{\ddot{\boldsymbol{q}}})\Big)\Big], \tag{3}$$

where $\boldsymbol{z} \sim \mathcal{N}(0,1)$ using randn function in Python with the seed set to 42. $D(\ddot{\boldsymbol{q}})$ and $D(\widetilde{\ddot{\boldsymbol{q}}})$ represent the discriminator's estimated probabilities for the real and generated acceleration data, respectively. The generated accelerations $\widetilde{\ddot{\boldsymbol{q}}}$ are computed via $\boldsymbol{f}(\boldsymbol{q}, \dot{\boldsymbol{q}}, \boldsymbol{M}_R, \boldsymbol{C}_R, \boldsymbol{K}_R, \boldsymbol{\kappa}_{nl}; \boldsymbol{\lambda})$, with $\boldsymbol{\lambda} = [k_1, \dots, k_5] = \boldsymbol{P}(\boldsymbol{z})$. The discriminator is trained to distinguish real (labeled 1) from generated (labeled 0) accelerations generated by the model, $\boldsymbol{f}$, using the standard two-sstep GAN approach [65].

The discriminator reaches its optimal loss when it is unable to differentiate between real and generated samples as:

$$\mathcal{L}_D = -\mathbb{E}_{\ddot{\boldsymbol{q}} \sim p(\ddot{\boldsymbol{q}})}[\log(0.5)] - \mathbb{E}_{\boldsymbol{z} \sim p(\boldsymbol{z})}[\log(1 - 0.5)] = \log(4) = 1.386. \tag{4}$$

Simultaneously, the parameter-generator network is trained to reduce the probability that the discriminator correctly classifies the generated data. The parameter-generator loss consists of two components:

$$\mathcal{L}_P = \mathcal{L}_{adv} + \gamma \mathcal{L}_{\mathrm{MSE}}, \tag{5a}$$

$$\mathcal{L}_P = -\mathbb{E}_{\boldsymbol{z} \sim p(\boldsymbol{z})}\Big[\log\Big(D(\widetilde{\ddot{\boldsymbol{q}}})\Big)\Big] + \gamma \mathbb{E}_{\ddot{\boldsymbol{q}} \sim p(\ddot{\boldsymbol{q}})}\Big[\big\|\ddot{\boldsymbol{q}} - \widetilde{\ddot{\boldsymbol{q}}}\big\|^2\Big], \tag{5b}$$

where the first term is the adversarial loss, prompting the parameter-generator to produce realistic values, while the second term enforces similarity of the generated accelerations to the real ones via MSE. The hyperparameter $\gamma$ could be used to balance the importance of these two terms; however, we set it to 1 in this work to create equal importance between the two losses. We train the parameter-generator with the discriminator's parameters fixed, using target labels of 1 in $\mathcal{L}_{adv}$ to encourage the generation of parameters that make $\widetilde{\widetilde{q}}$ appear real to the discriminator.

*2.4.2. Uncertainty Quantification*

To assess parameter reliability, UQ is performed using the values obtained during training. After convergence, parameter estimates from each epoch are collected, and a normal distribution is fitted to each using MATALB®'s *fitdist* function. The corresponding probability density functions are then evaluated with the pdf function over $\pm 6$ standard deviations from the mean.

## 3. APPLICATION OF SIVA: MODEL UPDATING OF CANTILEVER BEAM WITH STRONGLY NONLINEAR ATTACHMENT

We employ data simulated from a similar cantilever beam with a strongly nonlinear end attachment presented in [66]. The beam is uniform, homogeneous, and made of steel (with modulus of elasticity $E = 180 \times 10^9$ N/m³, and density $\rho = 7800$ kg/m³), with a length of 1.524 m, a width of 0.0381 m, and a thickness of 0.0064 m. The beam is modeled using 15 Euler-Bernoulli beam elements and Table 1 presents the natural frequencies and damping ratios for the first seven linear normal modes. The damping ratios were obtained from a comparable experimental system and are typical for a steel cantilever beam with these dimensions. A linear spring with stiffness $k_{lin} = 1.1 \times 10^4$ N/m and a nonlinear spring with force proportional to the displacement cubed and stiffness of $k_{nl} = 10^8$ N/m$^3$ were attached transversely in parallel at the free end of the beam. A lumped mass ($m = 0.0522$ kg) is also included at the tip to represent the additional mass introduced by coupling the attachment to the beam. A representative schematic of the beam in this configuration is shown in Fig. 2. An impulsive force in the form of a half-sine pulse with a 0.00635 s duration and an amplitude of 2 kN, is applied at the same location. The transient response of the beam is simulated using MATLAB's *ode45* with relative and absolute tolerances set to $10^{-8}$. We set the time duration to 4 s and the sampling rate to 2 kHz. Alternatively, Python's *solve_ivp* function can be used to simulate the system response; however, we observed that *ode45* in MATLAB is faster than *solve_ivp* for these simulations [27].

Figure 3 shows the simulated displacement time series of the tip, its continuous wavelet transform (CWT) spectrum (normalized to a maximum amplitude of 1) [66,67], and the Fourier spectrum computed using the fast Fourier transform. The validation time utilizes impulsive forces with amplitudes of 1 kN and 3 kN, and these results are also illustrated in Fig. 3 for comparison with the training response. Figure 4(a) shows the training process, where the initial fluctuations in the losses reflect the competition between the parameter-generator and discriminator. Around epoch 200, the losses attempt to stabilize, with the discriminator loss approaching the theoretical value of $\ln(4)$. This close convergence suggests that the parameter-extractor has learned to produce parameters such that the model outputs accelerations that effectively deceive the discriminator. The right panel shows the first term of Eq. (5) approaches roughly $\ln(4)/2$, with the MSE dropping to a relatively small value, implying strong agreement between the simulated and generated accelerations.

In the following, we examine two approaches to obtain the model parameters:

- **Approach I**. Following training, from the parameter-generator, each parameter is sampled 1000 times [25]. The final model parameters are computed as the mean of these samples and are reported in Table 2.

- **Approach II**. Once convergence is achieved, training continues and the parameter values from each subsequent epoch are recorded until the target number of epochs is reached. The final model parameters are obtained by averaging these values. As shown in Fig. 4(b), this approach is applied from epoch 300 onwards, and the result is presented in Table 2. Since the parameter- generator is fed with new noise data, this approach illustrates the method's robustness to overtraining [27]. Nonetheless, in practice, Approach I is more efficient unless the methodology is continuously updated with new data.

If computational resources permit, the optimal parameter set can be selected by simulating the response using ode45 or similar to produce the displacements, $\widetilde{\boldsymbol{q}}_{tr}$, and comparing them with the true displacements, $\boldsymbol{q}_{tr}$. This selection minimizes the MSE between the two and can be done using identification data, validation data, or a combination of both [27]. Moreover, the resulting values of these approaches can serve as the starting points for optimization algorithms that adjust the parameters to match the measured time series, as demonstrated in [19,20,68,69].

UQ is possible for both approaches through the process of fitting a distribution to the collected parameter value samples. To demonstrate UQ, a normal distribution is fitted to the collected parameters obtained in Approach II. Note that this could also be achieved using the parameters from Approach I. The distributions shown in blue in Fig. 4(c) reveal that the estimated parameters cluster closely around their respective mean values with small standard deviations, suggesting reliable parameter estimation. In each subplot, the dashed red line marks the mean value, the green-shaded region shows the 95% confidence interval (CI), and the exact value appears as a solid black line.

For comparison, we present the values identified by the SINDy method [44] in Table 3. For the candidate functions, we used the velocities, displacements, and the displacement of the tip raised to the powers of two through five. To evaluate the accuracy of the identification, the MSE is computed between the exact displacement at the tip, $q_N$, and the simulated time series signal, $\tilde{q}_N$ using each set of identified parameters. As can be seen in Table 3, SIVA produces accurate parameters, while SINDy struggles in this scenario. It is worth noting that the calculation time used by the SIVA for Approach II is 2 hr 5 min for 1000 epochs, whereas SINDy's output is obtained in a fraction of a second. Nonetheless, SIVA incorporates validation and UQ, as well as can includes additional datasets for both identification and validation cases. The calculations were performed on a PC with an Intel Core i7-8700 CPU @ 3.2 GHz, 32 GB RAM, and Windows 10 64-bit operating system.

Figure 5(a) presents the results from Approach I for the training cases (impact of 2 kN). The 95% confidence interval generated using 1000 parameter values is shown as the pink shaded region and the mean of this interval is provided as the blue line. The green line shows the simulated signal using the mean values calculated from the 1000 parameter sets. When comparing the displacement time series and Fourier spectra between the exact system and the responses from Approach I, the nonlinearities are accurately reproduced, and the narrow confidence bounds demonstrate minimal variation of the sampled parameters. Employing the parameters identified from Approach II listed in Table 3, we simulated the system response using identical initial conditions and force to those used for the exact system. Figure 5(b) shows that the comparison of the displacement at the tip for the exact system and the identified model, which shows that the identified model exhibits strong agreement with the exact response. This outcome verifies the ability of the method to accurately identify the unknown parameters of this highly nonlinear system. Considering that the system has no external factors or the presence of noise, and the exact signal was part of the identification process, close agreement was anticipated.

To further validate the proposed method, the exact and identified systems were simulated for impacts of 1 kN and 3 kN using the model identified using Approach I, which provides comparison with the validation datasets. These results are presented in Figs. 6(a) and (b) for 1 kN and 3 kN, respectively, and include the displacement time series of the tip, the corresponding CWT spectra, and the Fourier spectra. A strong agreement is observed between all three representations of the results. These results demonstrate that either method can accurately identify unknown parameters using existing partial knowledge obtained from a tuned linear model (FE model in this example), such that that existing model can be updated to capture additional linearities and nonlinearities added to the base system.

**4. Conclusions**

Different than traditional system identification methods, the system identification via validation and adaptation (SIVA) method incorporates simultaneous model validation through an independent validation dataset and provides inherent uncertainty quantification as well. SIVA employs a parameter-generator network that transforms random noise into physically meaningful parameter values. The process is guided by an adversarial loss that assesses how effectively the model-generated accelerations for validation datasets match their real equivalents, combined with the mean square error between the real accelerations and those generated by the identified model for the identification dataset. Furthermore, uncertainty quantification is performed on the collected set of parameters.

In this work, we applied SIVA to the nonlinear model updating problem where an existing linear model is updated to incorporate linearities and nonlinearities introduced by some addition to the underlying linear system. Furthermore, model reduction was employed to produce reduced mass, damping, and stiffness matrices with DOFs that correspond to measurable positions, such that the method could be applied to experimental data. The reduced matrices, transient vibration responses, and a reduced-order mathematical model of the beam were employed to identify the linear and nonlinear stiffnesses present at the tip of a beam, introduced by a nonlinear attachment to the underlying cantilever beam. The results indicate that SIVA can perform precise parametric system identification for highly nonlinear systems.

While SIVA effectively identifies system parameters, it assumes knowledge of the governing equations of motion, and future work will focus on the generalization of this assumption. Its performance remains acceptable with additional terms, but it degrades if essential physics are missing from the proposed model. Additional future work includes extending SIVA to partially or fully unknown governing equations. More importantly, future investigations will be focused on applying SIVA for experimental scenarios where a suitable dimensionality reduction method needs to be applied, and noise is a concern and should be filtered out or smoothed as effectively as possible. Finally, since SIVA is based on the original GAN, known for its instability [70], and incorporating advanced GAN variants could improve both accuracy and efficiency.

**FUNDING**

This research was supported by the Air Force Office of Scientific Research (AFOSR) Young Investigator Program (YIP) under grant number FA9550-22-1-0295.

**NOMENCLATURE**

| | |
|---|---|
| *CI* | Confidence interval |
| CWT | Continuous wavelet transform |
| DOF | Degree of freedom |
| *FE* | Finite element |
| GAN | Generative adversarial network |
| MSE | Mean square error |
| *NN* | Neural network |
| SEREP | System equivalent reduction expansion process |
| *SI* | System identification |
| SINDy | Sparse identification of nonlinear dynamics |
| SIVA | System identification via validation and adaptation |
| *UQ* | Uncertainty quantification |
| $\boldsymbol{f}(\boldsymbol{q}, \dot{\boldsymbol{q}}, \boldsymbol{M}_R, \boldsymbol{C}_R, \boldsymbol{K}_R, \boldsymbol{\kappa}_{nl}; \boldsymbol{\lambda})$ | Proposed reduced-order model |
| $\boldsymbol{M}_R$ | Linear mass matrix reduced from a full model |
| $\boldsymbol{C}_R$ | Linear damping matrix reduced from a full model |
| $\boldsymbol{K}_R$ | Linear stiffness matrix reduced from a full model |
| $\boldsymbol{q}$ | Displacement vector |
| $\dot{\boldsymbol{q}}$ | Velocity vector |
| $\ddot{\boldsymbol{q}}$ | Acceleration vector |
| $\boldsymbol{\kappa}_{nl}$ | Model nonlinear force vector |
| $\boldsymbol{\lambda}$ | Unknown parameters representing the addition to the system |
| $\boldsymbol{q}_{tr}$ | Training displacement vector |
| $\dot{\boldsymbol{q}}_{tr}$ | Training velocity vector |
| $\ddot{\boldsymbol{q}}_{tr}$ | Training acceleration vector |
| $\boldsymbol{q}_{val}$ | Validation displacement vector |
| $\dot{\boldsymbol{q}}_{val}$ | Validation velocity vector |
| $\ddot{\boldsymbol{q}}_{val}$ | Validation acceleration vector |
| $P$ | Parameter-generator network |
| $z$ | Random noise chosen with mean of 0 and standard deviation of 1 |
| $D$ | Discriminator network |
| $\tilde{\ddot{\boldsymbol{q}}}_{tr}$ | Model-generated acceleration during training |
| $\tilde{\ddot{\boldsymbol{q}}}_{val}$ | Model-generated acceleration for validation |
| $\tilde{\ddot{\boldsymbol{q}}}$ | Fake accelerations |
| $\boldsymbol{F}(t)$ | External forcing vector |
| $\boldsymbol{\kappa}_{nl}(\boldsymbol{q})$ | Model for the nonlinear element acting at beam tip |
| $k_{lin}$ | The exact linear stiffness introduced at the beam tip |
| $k_{nl}$ | The exact nonlinear stiffness introduced at the beam tip |
| $\mathcal{L}_D$ | Discriminator loss |
| $D(\ddot{\boldsymbol{q}})$ | Estimated probabilities for real acceleration data |
| $D(\tilde{\ddot{\boldsymbol{q}}})$ | Estimated probabilities for generated acceleration data |
| $\mathcal{L}_P$ | Parameter-generator loss |
| $\mathcal{L}_{adv}$ | Adversarial loss |
| $\mathcal{L}_{MSE}$ | MSE similarity loss |
| $\gamma$ | Hyperparameter balancing the importance of adversarial and MSE loss |
| $q_N$ | Exact displacement of the beam tip |
| $\tilde{q}_N$ | Simulated displacement of the beam tip |

**Figure Captions List**

Fig. 1    Flowchart of the proposed SIVA method for parametric system identification

Fig. 2    Schematic of the cantilever beam with a nonlinear spring used in the analytical study

Fig. 3    Simulated displacement response of the tip

Fig. 4    Training dynamics of the proposed SIVA: (a) Losses, (b) Identified parameters distributions of the end attachment, and (c) the distributions of the parameters

Fig. 5    Comparison of the displacement responses and Fourier spectra for the exact system and the identified model for an impact of 2 kN (the training case). Simulation of signals from parameters obtained: (a) Approach I and (b) Approach II.

Fig. 6    Comparison of the exact and predicted responses using the model from Approach I with impact forces of (a) 1 kN and (b) 3 kN

**Table Caption List**

Table 1      Architecture details of the parameter-generator and discriminator networks

Table 2      Frequencies and viscous damping ratios for the first 7 linear normal modes

Table 3      Comparison of the coefficients of the beam model

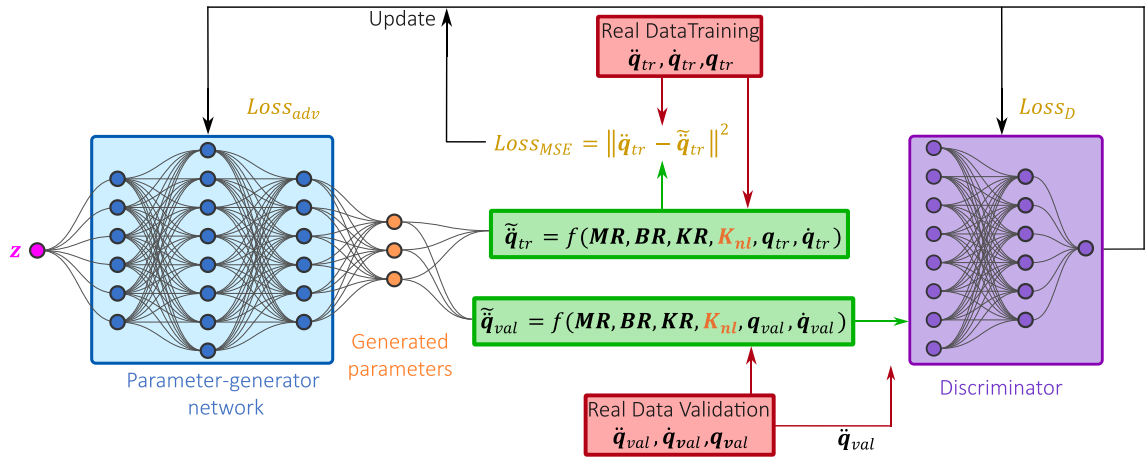

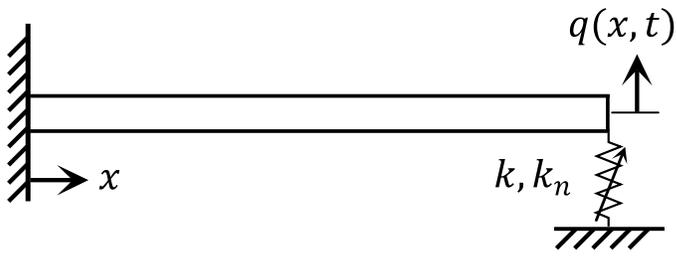

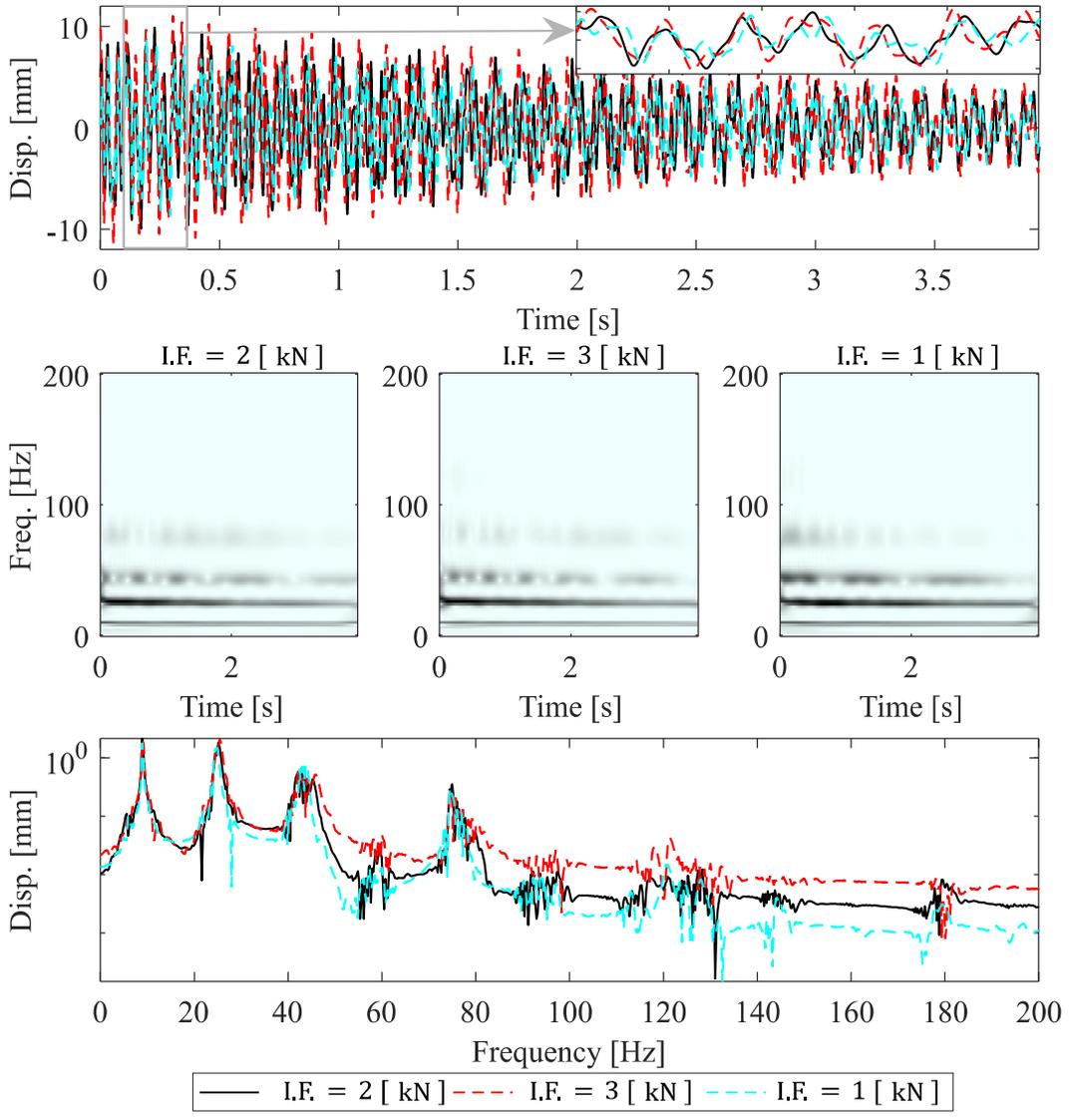

(a)

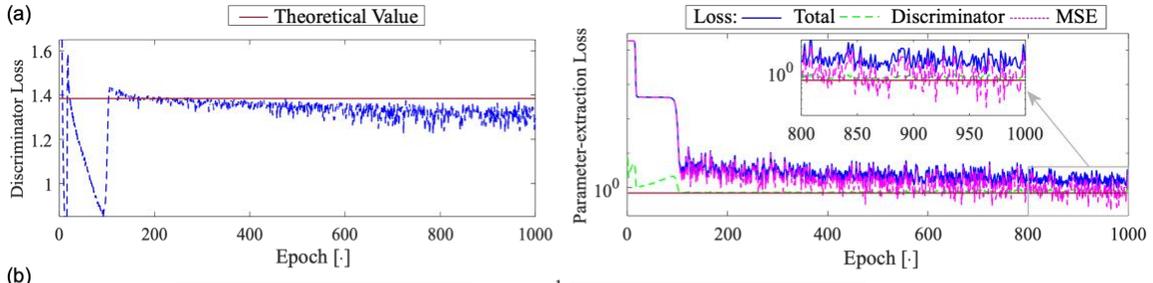

(b)

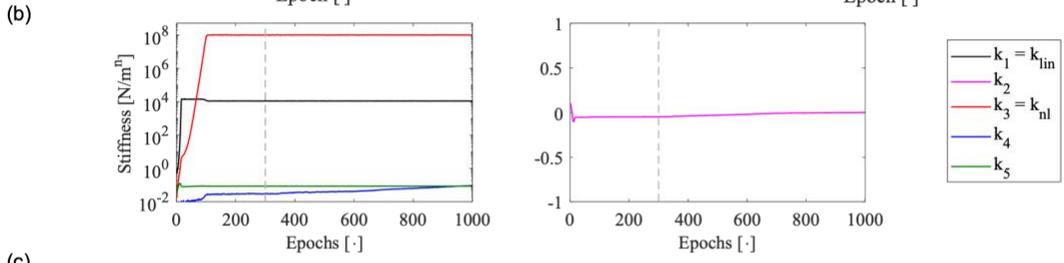

(c)

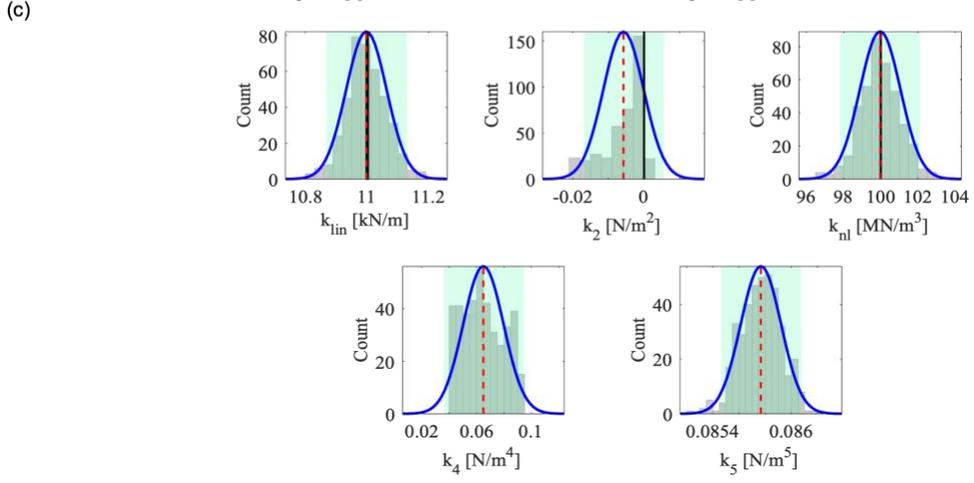

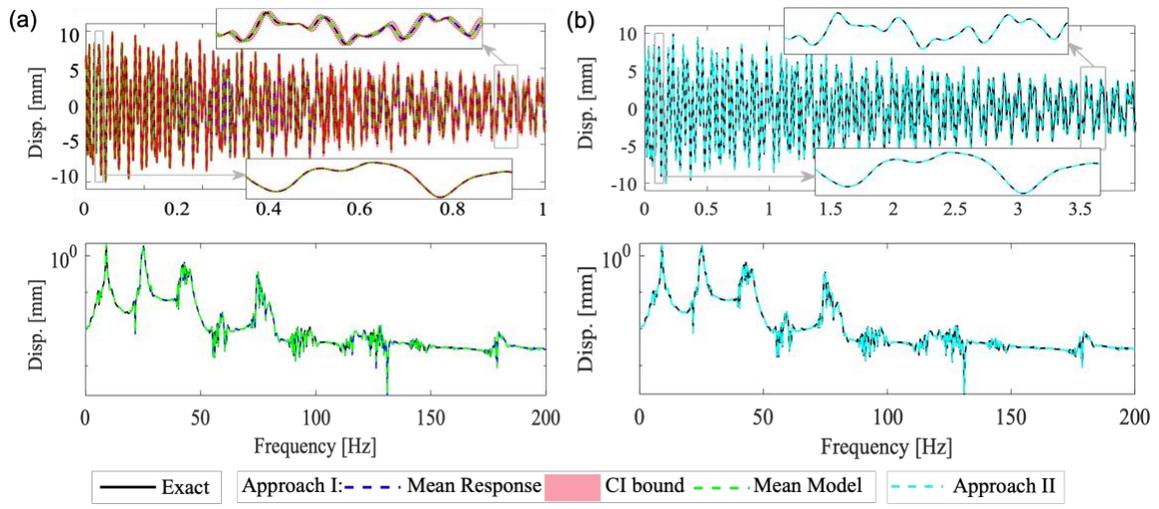

(a)                                              (b)

Exact system    Identified Model        Exact system    Identified Model

——— Exact system    - - - Identified Model

| Feature | Parameter-generator | Discriminator |
|---|---|---|
| Input | Random noise vector $\boldsymbol{z}$ | Acceleration signals |
| Layers (fully connected) | $64 \rightarrow 32 \rightarrow 16 \rightarrow n$ neurons | $64 \rightarrow 32 \rightarrow 1$ neurons |
| Activation functions | `LeakyReLU` (slope = 0.2, on hidden layers) `Linear` (output layer) | `LeakyReLU` (slope = 0.2, on hidden layers) `Sigmoid` (output layer) |
| Output | Parameters $\boldsymbol{\lambda}$: $\mathbf{a}$, $\mathbf{b}$ | Scalar (probability) |
| Special feature | Apply scientific notation: $k = \mathrm{a} \cdot 10^{\mathrm{b}}$ | — |

| Mode # | Frequency [Hz] | Damping ratio |
|--------|----------------|---------------|
| 1 | 2.079 | 0.0069 |
| 2 | 13.05 | 0.0052 |
| 3 | 36.61 | 0.0014 |
| 4 | 71.86 | 0.0017 |
| 5 | 119.0 | 0.0044 |
| 6 | 178.1 | 0.0038 |
| 7 | 249.4 | 0.0042 |

| Coefficient | Exact | SINDy | SIVA | |
|---|---|---|---|---|
| | | | Approach I | Approach II |
| $k_1 = k_{lin}$ [N/m] | $1.1 \times 10^4$ | $1.1734 \times 10^5$ | 11000 | 10999 |
| $k_2$ [N/m$^2$] | 0 | $-13.342$ | $3.4660 \times 10^{-4}$ | $-1.7883 \times 10^{-2}$ |
| $k_3 = k_{nl}$ [N/m$^3$] | $1 \times 10^8$ | $7.8052 \times 10^8$ | $9.9941 \times 10^7$ | $9.9974 \times 10^7$ |
| $k_4$ [N/m$^4$] | 0 | $7.4634 \times 10^5$ | 0.092635 | 0.052873 |
| $k_5$ [N/m$^5$] | 0 | $-2.7386 \times 10^9$ | 0.085777 | 0.052873 |
| MSE | 0 | $1.00 \times 10^2$ | $2.46 \times 10^{-4}$ | $1.02 \times 10^{-3}$ |